# Investigating the relationship between bolide entry angle and apparent direction of infrasound signal arrivals

Elizabeth A. Silber[1]

[1]Geophysics, Sandia National Laboratories, Albuquerque, NM, 87123 (esilbe [at] sandia.gov)

**Abstract**

Infrasound sensing offers critical capabilities for detecting and geolocating bolide events globally. However, the observed back azimuths, directions from which infrasound signals arrive at stations, often differ from the theoretical expectations based on the bolide's peak brightness location. For objects with shallow entry angles, which traverse longer atmospheric paths, acoustic energy may be emitted from multiple points along the trajectory, leading to substantial variability in back azimuth residuals. This study investigates how the entry angle of energetic bolides affects the back azimuth deviations, independent of extrinsic factors such as atmospheric propagation, station noise, and signal processing methodologies. A theoretical framework, the Bolide Infrasound Back-Azimuth EXplorer Model (BIBEX-M), was developed to compute predicted back azimuths solely from geometric considerations. The model quantifies how these residuals vary as a function of source-to-receiver distance, revealing that bolides entering at shallow angles, e.g., 10°, can produce average residuals of 20°, with deviations reaching up to 46° at distances below 1000 km, and remaining significant even at 5,000 km (up to 8°). In contrast, bolides with steeper entry angles, e.g., >60°, show smaller deviations, typically under 5° at 1000 km and diminishing to <1° beyond 5000 km. These findings attest of the need for careful interpretation when evaluating signal detections and estimating bolide locations. This work is not only pertinent to bolides but also to other high-energy, extended-duration atmospheric phenomena such as space debris and reentry events, where similar geometric considerations can influence infrasound arrival directions.

## 1. Introduction

Infrasonic detection and the study of asteroids and large meteoroids impacting the Earth's atmosphere fundamentally rely on the shock phenomena produced during their hypersonic entry. Asteroid entries into the Earth's atmosphere are accompanied by the formation of a strong shock wave (Bronshten, 1983; Silber et al., 2018), whose effects could pose a significant risk to life, infrastructure, and the environment (Chapman, 2008; Harris and D'Abramo, 2015; Toon et al., 1997; Trigo-Rodríguez, 2022). These shock waves, especially from larger asteroids, have the potential to cause severe damage on the ground due to direct impacts or lower atmospheric airbursts, as demonstrated by the Chelyabinsk bolide (Le Pichon et al., 2013; Popova et al., 2013). Currently, if an asteroid exceeds certain critical thresholds in size, composition, and velocity, the options for mitigating the consequences of its direct impact or low atmospheric airburst, are limited (e.g., Bender et al., 1995; Trigo-Rodríguez, 2022). Although efforts are underway to catalog Near-Earth Objects (NEOs) that might be hazardous, this task remains incomplete. Particularly challenging is the detection of NEOs ranging from meters to tens of meters in size, which are numerous but faint and difficult to identify using optical methods alone (Boslough et al., 2015). Detecting asteroids *a priori*, i.e., just before they enter the Earth's atmosphere, is theoretically possible but extremely challenging (e.g., Jenniskens et al., 2021; Jenniskens et al., 2009). Current observations indicate that the lead time for detection of an imminent impact is typically very short, which limits the ability to organize extensive observational efforts (Silber et al., 2024). Asteroids approaching from the direction of the Sun are particularly challenging to detect and may impact without prior warning. The Chelyabinsk bolide, whose shock wave caused injuries and significant damage (Popova et al., 2013), exemplifies this issue. Thus, enhancing the detection and characterization of asteroidal impacts is vital for understanding impact rates, refining observational techniques, and developing effective mitigation strategies.

The expansion of all-sky camera networks worldwide has improved our ability to monitor the night sky, but their coverage is still confined to specific geographic regions (Colas et al., 2020; Devillepoix et al., 2020). While space-based observations can offer global coverage (Jenniskens et al., 2018; Nemtchinov et al., 1997), no dedicated system exists for tracking bolides specifically. Currently, only U.S. Government (USG) sensors provide global detection capabilities, identifying and reporting bright flashes from bolides (Nemtchinov et al., 1997). However, these sensors are not exclusively designed for bolide detection. As a result, there is a notable lack of dedicated global systems for comprehensive bolide monitoring and analysis. Therefore, expanding infrasound-based detection and analysis of bolide events is especially valuable for advancing global threat assessments, as it complements optical techniques and can help identify smaller or more elusive near-Earth objects that frequently escape conventional surveys.

Extraterrestrial objects enter the atmosphere at hypersonic speeds, 11.2 – 72.8 km/s (Ceplecha et al., 1998) and subsequently generate shock waves that eventually decay to low frequency (<20 Hz) sound waves or infrasound (Silber et al., 2018; Tsikulin, 1970). Depending on the source parameters (e.g., size, velocity) as well as atmospheric conditions along the propagation path, infrasonic waves could be detected by microbarometers at large distances (100s to 1000s kilometers) (ReVelle, 1976). Microbarometers record minute variations in atmospheric pressure as a function of time, capturing these as a waveform or timeseries. Analyzing timeseries can yield valuable information about the source, including apparent direction of infrasound wave arrival at the station (back azimuth), signal period and amplitude. The analysis of back azimuth from two or more infrasound stations can be leveraged for estimating event geolocation. The signal period and amplitude can be used to estimate energy deposition by the asteroid (for further details and the list of energy relations, see Silber and Brown (2019), Ens et al. (2012), Gi and Brown (2017)).

Over the last few decades, infrasound sensing has emerged as a critical tool for detecting, locating, and characterizing asteroid entries around the world (e.g., Le Pichon et al., 2013; Ott et al., 2019; Pilger et al., 2019; Silber, 2024b). Its continuous and global monitoring capabilities offer an essential complement to other detection methods, improving their effectiveness and providing independent observations when ground truth is limited. Unlike optical methods, infrasound operates independently of daylight and weather conditions, making it particularly valuable for detecting bolides at all times and under various atmospheric conditions. This robustness allows infrasound to capture events that may elude optical sensors, thus contributing to a more comprehensive understanding of bolide impacts and their consequences (e.g., Arrowsmith et al., 2008; Ott et al., 2019; Pilger et al., 2020; Silber et al., 2011; Wilson et al., 2025). Therefore, infrasound, through its global reach and long-range detection capabilities, plays a critical role in evaluating bolide impact energy and refining influx rate estimates. Understanding these parameters is fundamental for assessing the risks associated with future impacts, guiding the development of mitigation strategies to protect infrastructure and populations.

From the moment bolides start generating shock waves at a high altitude (typically >75 km), they deposit energy over their entire trajectory, which could present unique challenges for interpreting infrasound signals. The detectability of these signals is influenced by several factors, including the type of shock wave (cylindrical or spherical) generated and the trajectory's orientation relative to monitoring stations (Silber, 2024a; Wilson et al., 2025). Bolide trajectories can vary significantly in length, ranging from tens to hundreds of kilometers, and are affected by the entry angle (e.g., Moreno et al., 2016; Shober et al., 2020). This variability impacts how signals are detected and localized, especially at close ranges. Predicted back azimuths (the apparent direction of infrasound signal arrival at the station) and signal travel times can vary significantly due to the bolide's trajectory in respect to the station, as well as its distance (Silber, 2024a). A notable case illustrating the impact of trajectory geometry on infrasound detection is the recent observation of the National Aeronautics and Space Administration (NASA) Origins, Spectral Interpretation, Resource Identification, and Security–Regolith Explorer (OSIRIS-REx) sample return capsule (Lauretta et al., 2017) during its re-entry (Silber and Bowman, 2025; Silber et al., 2024). It is generally assumed that at very large distances, the trajectory may be approximated as a point source because its length becomes negligible compared to the distance from the source to the station (ReVelle, 1976). However, this supposition may not always be valid, particularly for energetic events that emit directional acoustic energy (Pilger et al., 2015), as will be discussed later.

Infrasound is subject to propagation effects (e.g., winds, small scale perturbations, turbulence, travel through different atmospheric waveguides) along its path from source to receiver. Thus, it is expected that signals might exhibit some small deviation from true back azimuth, which can be calculated based on known ground truth location and infrasound station location. The latter is always known, and ideally, the former should come from well-constrained observations, such as USG sensors or optical measurements. Oftentimes, however, ground truth might be absent, poorly defined, or require further validation. In such a case, multi-station detections of infrasound signals from a given event could provide useful constraints for geolocation, provided that the back azimuths are accurately determined.

As it was pointed out in a recent review paper (Silber, 2024a), the degree of variability in observed versus true back azimuths varies among bolide events and this could be, in part, related to the bolide's trajectory geometry. In fact, some bolide infrasound detections show observed back azimuth variability that surpasses values normally anticipated from stationary sources (typically

<10°). If we set aside propagation effects, station noise, and signal analysis techniques, a critical question arises: to what extent can variability in back azimuths be ascribed to the bolide's trajectory geometry alone? Additionally, in the absence of definitive ground truth, it is valuable to determine whether an uncertainty range for the estimated location can be established based on the expected variability in back azimuths due to the bolide's trajectory. Furthermore, it is useful to estimate the distance from the source at which the largest deviations from the true back azimuth might occur. This helps in assessing whether detections beyond a certain value of back azimuth are feasible due to bolide geometry alone, and in determining whether such deviations at large distances should be considered anomalies or expected variability.

Infrasound analysis has traditionally been conducted under the assumption that sources are stationary point sources, a premise well suited for non-proliferation monitoring. However, with increasing interest in bolides and a growing number of orbital debris and reentry events, there is a pressing need to rigorously quantify the geometric effects on infrasound signal arrival directions. This study addresses these issues by first examining observed back azimuth deviations in highly energetic bolides, thereby illustrating the prevalence and magnitude of such phenomena in real-world data. These observations set the stage for a subsequent theoretical investigation, in which a geometric framework is developed to quantify the maximum possible back azimuth variability as a function of the bolide's (or any similarly extended object's) entry angle and the distance from the source to the receiving station. By separating geometric considerations from extrinsic factors (e.g., atmospheric propagation, monitoring station effects), this study provides a rigorous baseline for understanding how the orientation and length of a bolide trajectory alone can result in significant deviations from the nominal (point-source) arrival direction. In the observational portion, emphasis is placed on the most energetic and well-constrained events, since they generate sustained shock production over a substantial vertical column, rendering them particularly amenable to probing geometry-induced deviations. Although the present study is demonstrated using bolide events, the underlying principles and resultant bounds on azimuthal deviations are broadly applicable to other high-energy, prolonged atmospheric phenomena (e.g., the recent OSIRIS-REx sample return mission (Silber et al., 2024)). This paper is organized as follows: Section 2 describes the methodology, encompassing both the observed bolides dataset and the development of the theoretical model, Section 3 presents the results and discussion, and Section 4 outlines the conclusions.

## 2. Methods

## 2.1 Observed Bolides

### 2.1.1 Dataset

While our planet is impacted by extraterrestrial material on a daily basis, there are only a handful of very energetic events with accurate ground truth and accompanied infrasound detections. For the observational part, this study focuses on well-characterized, large bolides for which infrasound detections have been verified and published (Ens et al., 2012; Gi and Brown, 2017; Ott et al., 2019; Pilger et al., 2020; Silber et al., 2011). The infrasound portion of the dataset comprises back azimuth values calculated and reported in these earlier studies using established signal processing techniques (e.g., beamforming). Readers interested in the detailed infrasound signal processing methods are referred to Silber (2024a), which provides a comprehensive overview of these standard procedures. All events analyzed here (Ens et al., 2012; Gi and Brown, 2017; Ott et al., 2019; Pilger et al., 2020; Silber et al., 2011) were detected on well-calibrated stations of the International Monitoring System (IMS) of the Preparatory Commission of the Comprehensive Nuclear-Test-Ban Treaty (CTBT PrepCom) (Brachet et al., 2010; Christie and Campus, 2010). The IMS infrasound network currently includes 53 certified stations (out of a planned 60), capable of detecting any 1 kt explosion worldwide (National Research Council, 2012) (Figure 1), and has proven effective at monitoring bolide impacts on a global scale.

Ens et al. (2012) performed statistical analysis of infrasound signals, and identified 71 bolide detections on 143 stations. Gi and Brown (2017) expanded that list by finding additional detections for a total of 128 events on 267 stations. Some events were detected by only a single station, while others generated signals at two or more infrasound stations. More recently Ott et al. (2019) and Pilger et al. (2020) published additional and some overlapping events detected on the IMS stations. A key prerequisite for this exploratory study is a reasonably high energy deposition, thus only events with significant energy are selected. Energetic events are produced by large objects and there is a higher probability of a detection by multiple infrasound stations globally. Furthermore, energetic events are more likely to generate strong shock waves along their entire path through dense regions of the atmosphere, increasing the likelihood of detection of signals emanating from different parts of the trajectory at distant stations. Of course, this assumption does not account for detectability due to extrinsic factors, including station noise and propagation effects. Very low-energy events could potentially introduce biases into the results of this study.

Ground truth for these bolides was obtained from USG sensors, whose detections are archived and publicly reported by the NASA Jet Propulsion Laboratory (JPL) Center for Near-Earth Object Studies (CNEOS) webpage (https://cneos.jpl.nasa.gov/fireballs/). Since 1988, CNEOS has cataloged approximately 1000 bolides, listing parameters such as date, time, latitude, longitude, and energy released (total radiated energy and total impact energy), and, when available, peak brightness altitude and velocity vector. The velocity vector can be used to derive an entry angle and estimate orbital parameters (Peña-Asensio et al., 2022). Reported impact energies in this catalog range from 0.073 kt and to the 440 kt Chelyabinsk event (1 kt of TNT = 1.485E12 J). It is worth noting that recent studies (e.g., Hajduková et al., 2024) have highlighted that CNEOS estimates, particularly the velocity vector components, may exhibit significant biases and limited precision. In this study, however, the focus is exclusively on the relative differences between observed and predicted back azimuths (i.e., back azimuth residuals) rather than on absolute positional values; thus, any uncertainties in the CNEOS parameters do not affect the analysis.

The CNEOS database contains 13 events with an impact energy of 10 kt of TNT equivalent or greater. Of these, three events lack the velocity vector, and two additional events are not included in the published detections. One event was detected by only two stations, and since one of these stations provided an outlier value for the back azimuth, this event was excluded from the analysis. The remaining seven events that are examined in this work are listed in Table 1 and their locations are shown in Figure 1. These events have impact energies of 13 kt or greater and exhibit a range of entry angles, from shallow (16°) to steep (67°). The physical parameters are obtained from Peña-Asensio et al. (2022). Figure 2 contextualizes the impact velocity and peak brightness altitude of these events relative to the entire CNEOS bolide population.

### 2.1.2 Back Azimuth Residuals Analysis

A comparative analysis was performed to examine the back azimuth residuals (difference between the observed and predicted values) and to explore possible links to the entry angle and other parameters. In infrasound analysis, the back azimuth ($\alpha$) is typically defined as the compass bearing (measured clockwise from true north) at the receiving station indicating the incoming direction of the acoustic signal. This convention contrasts with the forward azimuth used in geodesy, where bearings are calculated from the perspective of a source. By taking the station as the initial point and the source as the final point, the calculated bearing naturally corresponds to the infrasound back azimuth. As described in Section 2.1, the observed back azimuth values were obtained from published studies (Ens et al., 2012; Gi and Brown, 2017; Ott et al., 2019; Pilger et al., 2020; Silber et al., 2011). Some events, notably the Indonesian bolide (8 October 2009) and the Chelyabinsk bolide (15 February 2013) were analyzed in different independent studies (Gi and Brown, 2017; Ott et al., 2019; Silber et al., 2009). Silber (2024a) noted that signal measurements, including observed back azimuths, might vary from study to study. Keeping note of this and to avoid duplication, only one dataset for a given bolide was selected for this work.

To determine the back azimuth residuals (i.e., the absolute difference between the observed and the predicted back azimuths), a Python code was developed to calculate the predicted back azimuth at each station. This approach uses the station's geographic coordinates and the bolide's peak brightness location (as reported by CNEOS), employing standard great circle bearing formula (Sinnott, 1984; Vincenty, 1975) to provide a robust comparison between observed and theoretical arrival directions (also see Silber (2024a)). The geopy Python library provides convenient functions (e.g., geopy.distance.geodesic or geopy.distance.great_circle) for calculating distances and bearings between latitude ($\phi$) and longitude ($\lambda$) pairs. However, in this work, the haversine and bearing formulae were implemented manually in Python for full control over the computations. The great circle distance $d$ (in km) between two latitude–longitude pairs ($\phi_1$, $\lambda_1$) and ($\phi_2$, $\lambda_2$) on Earth is given by the haversine formula:

$$d = 2R \, \sin^{-1}\left[\sqrt{\sin^2\left(\frac{\phi_2 - \phi_1}{2}\right) + \cos(\phi_1)\cos(\phi_2)\sin^2\left(\frac{\lambda_2 - \lambda_1}{2}\right)}\right] \qquad (1),$$

where $R$ = 63711 km (as per World Geodetic System 84 (WGS84), (USDMA, 1991)). The back azimuth ($\alpha$) is then obtained from:

$$\alpha = \tan^{-1} 2 \left[\sin(\lambda_2 - \lambda_1)\cos(\phi_2), \cos(\phi_1)\sin(\phi_2) - \sin(\phi_1)\cos(\phi_2)\cos(\lambda_2 - \lambda_1)\right] \quad (2).$$

For further details on the mathematical background, see Sinnott (1984) and Vincenty (1975).

**Table 1:** List of events selected for this exploratory study. The event metadata was extracted from the CNEOS database. The published infrasound detections are from Pilger et al. (2020) (a), Gi and Brown (2017) (b), and (Ott et al., 2019) (c). Entry angles come from Peña-Asensio et al. (2022). The number of infrasound stations that detected the event is also given.

| Date | Time [UTC] | Latitude [deg] | Longitude [deg] | Total radiated energy [J] | Impact energy [kt] | Entry angle [deg] | Altitude [km] | Velocity [km/s] | Number of detecting stations | * |
|------|-----------|----------------|-----------------|---------------------------|--------------------|--------------------|---------------|------------------|------------------------------|---|
| 18 Dec 2018 | 23:48:20 | 56.9 | 172.4 | 3.13E+13 | 49 | 68.6 | 26 | 13.6 | 20 | a |
| 6 Feb 2016 | 13:55:09 | -30.4 | -25.5 | 6.85E+12 | 13 | 22.1 | 31 | 15.6 | 3 | b |
| 15 Feb 2013 | 3:20:33 | 54.8 | 61.1 | 3.75E+14 | 440 | 15.9 | 23.3 | 18.6 | 14 | c |
| 25 Dec 2010 | 23:24:00 | 38 | 158 | 2.00E+13 | 33 | 60.8 | 26 | 18.1 | 9 | b |
| 21 Nov 2009 | 20:53:00 | -22 | 29.2 | 1.00E+13 | 18 | 28.6 | 38 | 32.1 | 3 | b |
| 8 Oct 2009 | 2:57:00 | -4.2 | 120.6 | 2.00E+13 | 33 | 67.4 | 19.1 | 19.2 | 11 | b |
| 7 Oct 2004 | 13:14:43 | -27.3 | 71.5 | 1.04E+13 | 18 | 27.3 | 35 | 19.2 | 4 | b |

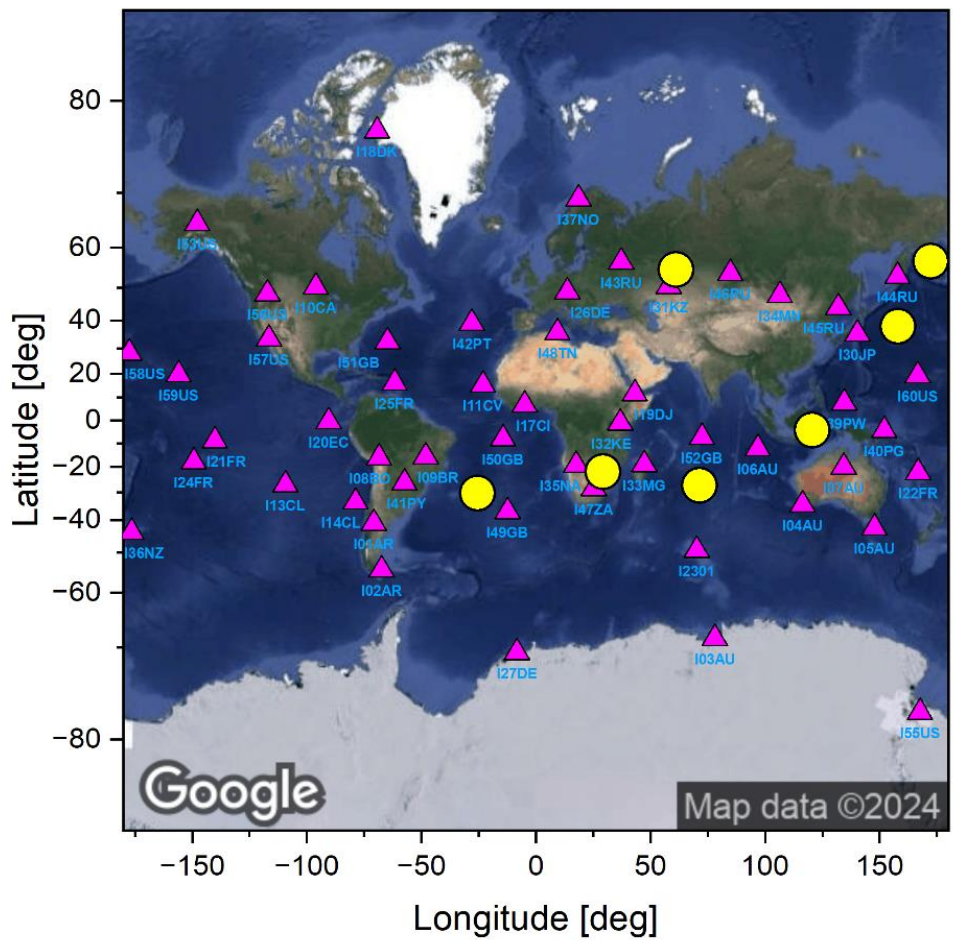

**Figure 1:** Map showing the currently operational infrasound stations of the IMS network and the energetic bolides from Table 1. As of writing this paper, there are 53 certified stations out of a planned 60.

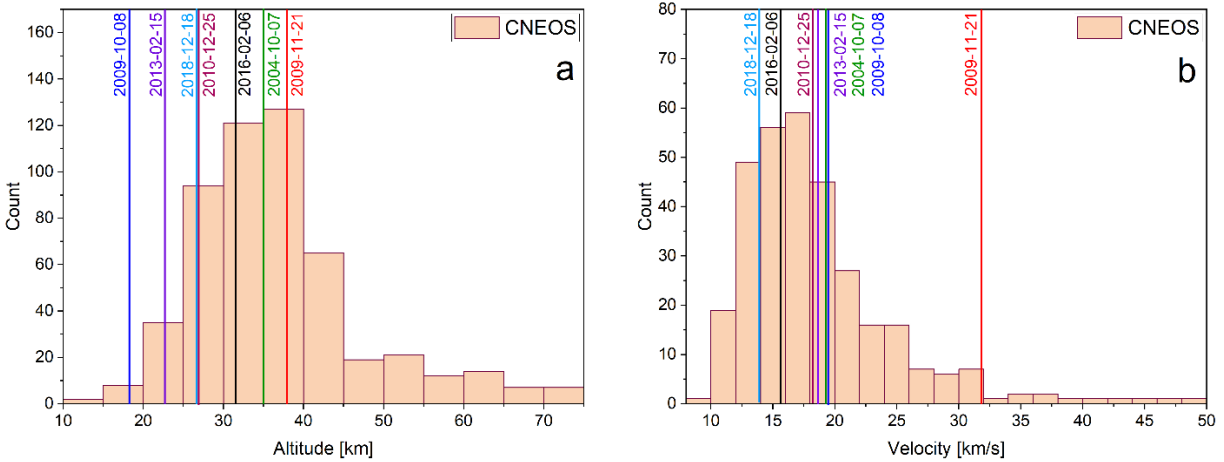

**Figure 2:** Histograms showing the entire CNEOS population of bolides that have their (a) peak brightness altitude (n = 532) and (b) velocity (n = 319) published. The events analyzed in this work are shown with color coded vertical lines.

### 2.3 Bolide Infrasound Back-Azimuth EXplorer Model (BIBEX-M)

Understanding the influence of bolide geometry on the apparent arrival directions of infrasound signals is critical for accurately interpreting these measurements. In practice, infrasound propagation is affected by numerous factors (e.g., atmospheric conditions, waveguides, noise). However, to rigorously quantify the purely geometric effect, independent of propagation complexities, a theoretical framework, hereafter referred to as the *Bolide Infrasound Back-Azimuth EXplorer Model* (BIBEX-M), was developed. BIBEX-M considers exclusively the geometry of the bolide trajectory, establishing a baseline against which full propagation models can later be compared.

In BIBEX-M, a hypothetical energetic bolide is assumed to produce shock waves continuously along its luminous path, yielding a ground-projected lateral distance (or ground track length, $L$) that depends on the entry angle $\theta$. Specifically:

$$L = (h_{start} - h_{end}) / \sin(\theta) \qquad (3),$$

where $h_{start}$ is the altitude at which the shock wave begins, and $h_{end}$ is the terminal altitude beyond which a shock wave is no longer generated. To cover a wide parameter space, BIBEX-M bounds the bolide's luminous path and the shock producing region between $h_{start} = 80$ km and $h_{end} = 10$ km. With these values, a bolide entering at an angle of 10° covers a lateral ground-projected distance of approximately 397 km, whereas one entering at 80° spans only 12.3 km. The choice of 80 km as the starting altitude and 10 km as the terminal altitude is based on typical altitudes where, in principle, bolide shock waves are effectively generated and detected (Ceplecha et al., 1998; Silber et al., 2018). It is important to note that these altitudes are not universal thresholds (e.g., Borovička et al., 2022; Devillepoix et al., 2019; Popova et al., 2013); rather, they represent an idealized scenario that captures the broadest plausible region of atmospheric shock production, thereby maximizing the potential for geometry-driven back azimuth deviations. By delineating this extended vertical range, BIBEX-M isolates the upper limit of the lateral ground track length and serves as a robust baseline for assessing how geometry alone can shape the apparent arrival directions in infrasound observations. The bolide's trajectory and ground track can differ substantially in both distance and orientation relative to each infrasound station. Consequently, acoustic signals emanating from distinct points along the trajectory, particularly at the starting and ending altitudes, may arrive at a station from markedly different directions. These boundary end-member points represent extreme cases within the parameter space, capturing the maximum possible variation in back azimuth attributable solely to geometric effects.

To capture a broad range of realistic scenarios, 200 seed points were randomly generated worldwide using a custom Python script (employing libraries such as NumPy and the random module). Each seed location is defined by a latitude ($\phi$) and longitude ($\lambda$) pair and stratified into four latitude bands to achieve a balanced distribution: equatorial ($\pm[0–9°N]$, 20 points), low-latitude ($\pm[10–44°N]$, 80 points), mid-latitude ($\pm[45°N]$, 20 points), and high-latitude ($\pm[46–80°N]$, 80 points) (Figure 3). To ensure spatial diversity, the random location generator was constrained such that no two seeds are separated by less than 2° in angular distance. For each seed point, trajectories were computed in eight directions (the four cardinal directions plus four intercardinal directions: NE, NW, SE, SW). The endpoints of each trajectory were then determined by translating the seed location along the specified bearing for the distance $L$, computed from the chosen entry angle $\theta$. This procedure yielded 27,200 distinct trajectories spanning entry angles from 5° to 85°, in 5° increments, corresponding to ground tracks lengths ranging from approximately 6.1 km to 800 km.

For every trajectory, the back azimuth at each International Monitoring System (IMS) infrasound station worldwide was computed from both the starting and ending points of the bolide's path.

Distances and azimuths between the station and each endpoint were calculated via standard geodesy formulae (Sinnott, 1984; Vincenty, 1975) (Eqs. (1-2)). For each infrasound station globally, the back azimuths and source-station distances corresponding to the starting and ending points of the ground track were computed. The total number of source-station pairs initially computed was $1.632 \times 10^6$. The back azimuth residual, defined as the difference between these back azimuth values from the two endpoints, reflects the maximum possible geometric deviation along the bolide's shock wave path as 'seen' by a station. To avoid complications arising from the Earth's curvature, only source-stations pairs with distances less than 15,000 km were retained, resulting in the final dataset of just over 1.2 million pairs.

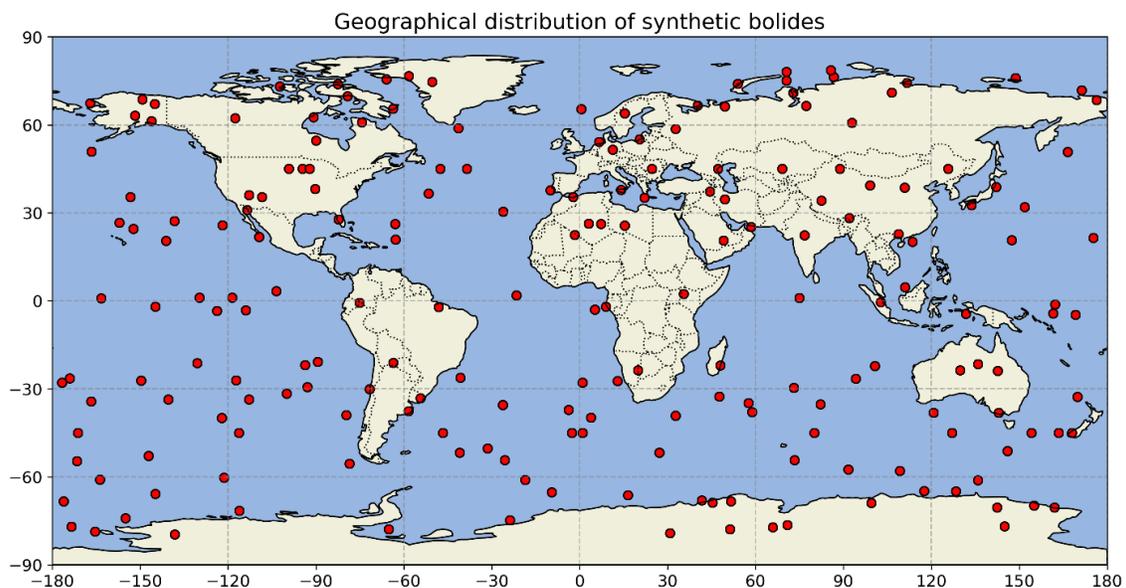

**Figure 3:** Distribution of synthetic bolide seed locations. Each point represents a randomly generated seed, stratified into four latitudinal bands to ensure uniform global coverage. From each seed, bolide trajectories are constructed for entry angles ranging from 5° to 85° (in 5° increments) and for eight different orientations (cardinal and intercardinal directions), providing a comprehensive framework for assessing the geometric influences on back azimuth deviations.

## 3. Results and Discussion

### 3.1 CNEOS Bolides

The back azimuth residuals for the seven events are shown in Figure 4. To more clearly depict the small residuals for the Bering Sea bolide (18 Dec 2018), panels (a) and (b) use different horizonal scales. Residuals vary across other events; for instance, the 2016 bolide has a residual of 25° at one station, and a similar residual is observed for the Indonesian bolide (8 October 2009). In contrast, the Chelyabinsk superbolide exhibits non-negligible residuals across all infrasound stations. This widespread deviation is particularly striking because the expected direction of infrasound wave arrival does not consistently align with the point of peak brightness reported on the CNEOS website. This could be, in part, due to the shallow entry angle (16°) and the high energy of the bolide.

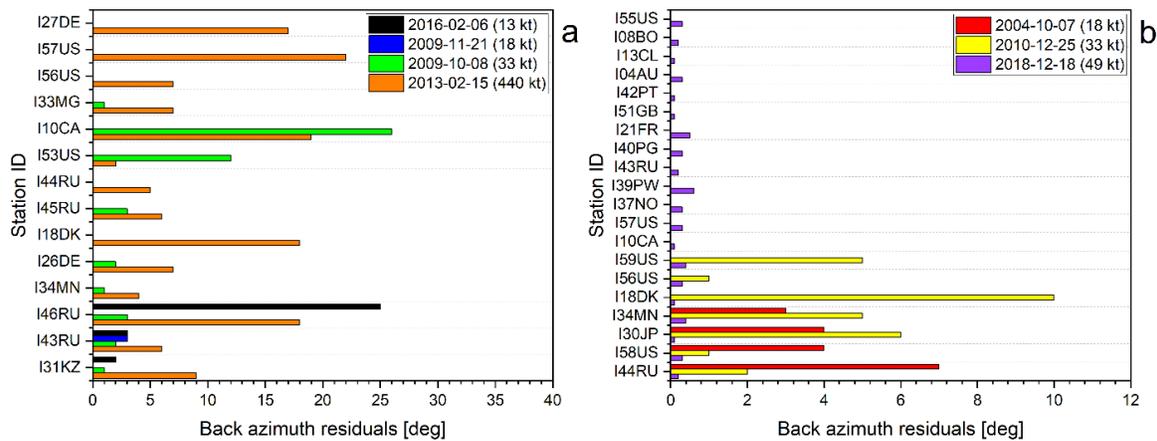

**Figure 4**: Back azimuth residuals for the seven events investigated here. Note that the horizontal scale is not the same in panels (a) and (b). The Chelyabinsk bolide (15 Feb 2013) has notable residuals across all infrasound stations, reflecting the effects of its shallow entry angle and high energy deposition on the shock wave propagation.

Before delving further into the problem, it is useful to visualize the global distribution of detections and the relative residuals for each event. Figure 5 displays maps for each event, indicating both the location of peak brightness as reported on the CNEOS website (see Table 1) and the positions of infrasound stations that recorded the event. The stations are color-coded based on their relative back azimuth residuals, with red representing the greatest deviation from the true back azimuth and deep green indicating the smallest deviation. The scale for back azimuth residuals varies between panels to accommodate the different ranges observed for each event. Infrasound stations that did not detect the event are shown in grey. Additionally, the bolide's azimuth, which represents the direction of its propagation path, is displayed in the upper right corner of each panel.

Light curves (Figure 6) have been digitized from files available on the CNEOS webpage and adjusted to reflect the reported peak brightness, and the start time of zero seconds. It is important to note that minor discrepancies may exist in the CNEOS database for some events, and future updates may address these issues. At present, it is unclear whether any of the light curves presented here are affected by such discrepancies. The altitude of peak brightness is annotated in each panel.

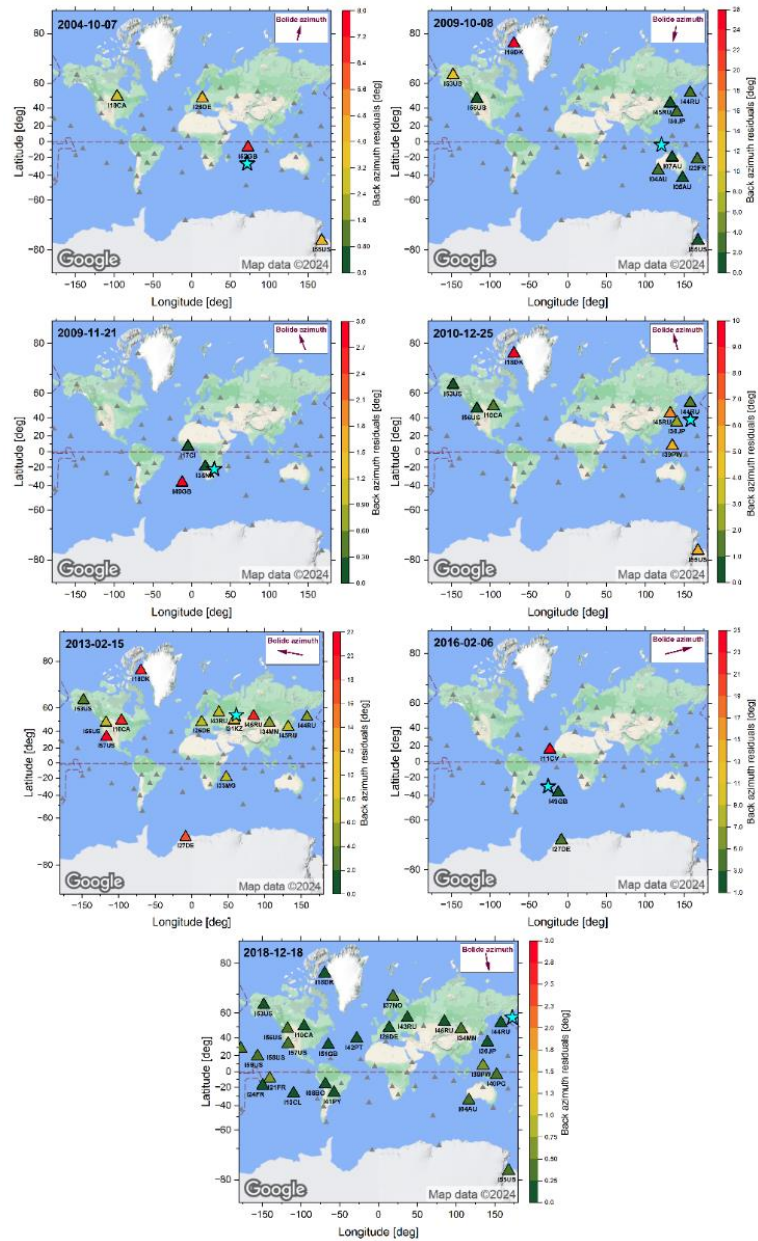

**Figure 5:** Map displaying the bolide peak brightness location (indicated by a star) along with the positions of IMS infrasound stations. Stations that detected the event are shown as colored triangles (with colors representing the magnitude of the back azimuth residuals), while those that did not detect the event are shown in grey. This spatial distribution provides essential context for evaluating the variability of back azimuth deviations relative to the bolide's reported position.

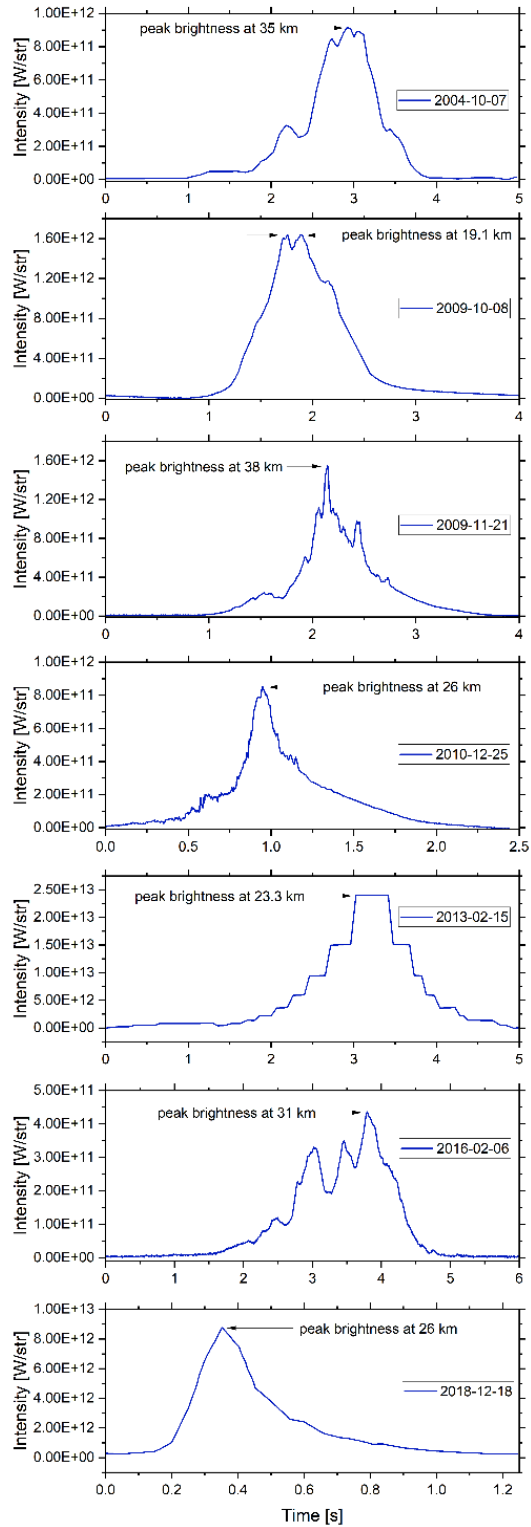

**Figure 6:** Light curves of bolide events with the altitude of peak brightness annotated. The light curves were digitized from files available on the CNEOS webpage.

Qualitatively, the light curves exhibit characteristics indicative of fragmentation (sudden peaks in brightness) and airbursts (a large peak followed by a rapid decline in brightness). For example, the 2016 bolide underwent multiple fragmentation episodes; in these scenarios, it is possible that different stations captured signals from different parts of the trail, resulting in deviations in the observed back azimuths relative to the single peak brightness point reported in the CNEOS database (also see Silber (2024a) for further discussion). It is important to recognize that the modes of shock production, such as a cylindrical line source or fragmentation episodes, can vary as a function of the bolide's flight path, with significant fragmentation typically occurring towards the end of the trail. Although a detailed discussion of propagation effects is beyond the scope of this paper, it is worth noting that atmospheric conditions along the propagation path can influence the transmission of infrasound. For instance, acoustic energy may be more effectively transmitted through stratospheric waveguides, such as the AtmoSOFAR channel (Albert et al., 2023), which can be leveraged for estimating the source altitude through propagation modeling (e.g., Silber, 2024b).

Although the analysis is limited by the very small sample size due to the scarcity of energetic events, a Spearman correlation (Spearman, 1904) matrix was computed to explore correlations among various parameters (Figure 7). The Spearman rank correlation coefficient is a nonparametric metric that evaluates how two variables co-vary in a monotonic manner by using ranked data, which makes it particularly robust in the presence of outliers or skewed distributions. A significance threshold ($p < 0.05$) was adopted to distinguish meaningful correlations from those that might arise by chance, given the small sample size. The parameters examined include back azimuth residuals, entry angle, peak brightness altitude, station-source distance, bolide diameter and bolide entry velocity. Notably, significant correlations were found between back azimuth residuals and both the entry angle ($p = -0.72$) and the bolide diameter ($p = 0.73$), suggesting that large bolides with shallower entry angles are more likely to exhibit higher back azimuth deviations. These correlations remain robust even when the Chelyabinsk bolide, an extreme case, is excluded from the analysis (entry angle $p = -0.70$, bolide diameter $p = 0.72$). Additionally, bolides with higher impact velocities also tend to display greater back azimuth residuals and shallower entry angles. While these initial findings are constrained by the limited number of events, they demonstrate the utility of Spearman correlation analysis in identifying key physical parameters that influence back azimuth variability. Future analyses with an expanded dataset will be necessary to confirm or refine the observations made in this study.

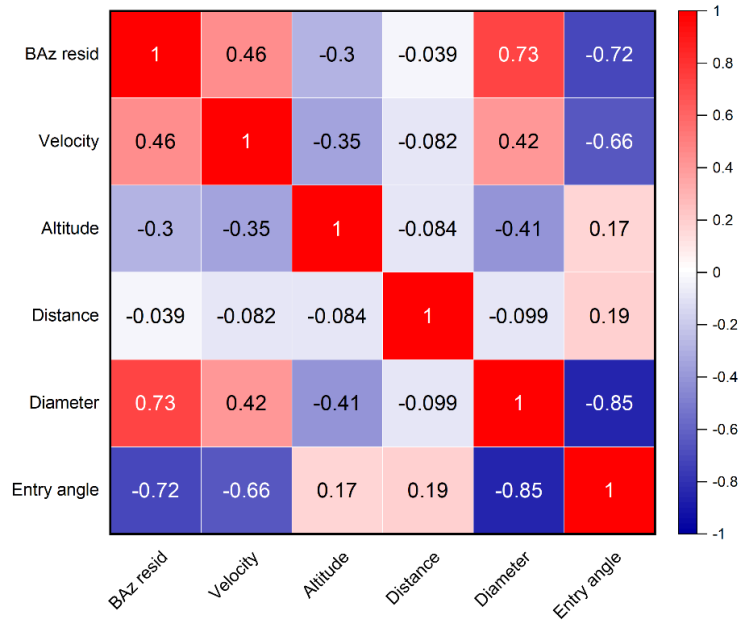

**Figure 7:** Spearman correlation matrix computed from the set of 7 energetic bolide events selected from the CNEOS database (see Table 1). The matrix examines the relationships among key parameters, including back azimuth residuals, entry angle, peak brightness altitude, station–source distance, bolide diameter, and bolide entry velocity, with notable correlations observed (e.g., a strong negative correlation between entry angle and back azimuth residuals, and a strong positive correlation between bolide diameter and back azimuth residuals).

Figure 8a-d shows panel plots with various quantities (a) peak brightness altitude, (b) bolide diameter, (c) impact velocity, and (d) source-station distance) versus back azimuth residuals. High-altitude events tend to exhibit more constrained back azimuths (Figure 8a), a pattern that may be influenced by limited sample sizes, as only three stations recorded each of the two highest altitude events. Signals from these high altitudes experience greater attenuation, and notably, these events had significantly lower energy levels (18 kt) compared to the Chelyabinsk event (440 kt). The Chelyabinsk asteroid is significantly larger than the rest of the impactors (Figure 8b). There seems to be a decrease in back azimuth residuals as the bolide size decreases (<8 m). In our analysis, the back azimuth residuals are controlled primarily by geometric factors, specifically, the length and orientation of the bolide's ground track, rather than by brightness or size per se. Any bolide observed at close range will indeed show significant residuals if its geometry results in a large angular separation between the beginning and end of its shock-producing path. However, in this data set, smaller objects tend to have steeper entry angles, which yield shorter ground tracks and, consequently, lower geometric deviations. Thus, while geometric effects dominate, the apparent trend of lower residuals for smaller objects can be explained by their association with steeper trajectories, rather than an intrinsic property of low energy or faintness. It is not clear why there is a bias towards smaller and steeper objects. This trend may suggest that smaller bodies entering at shallow angles are less likely to survive a prolonged atmospheric passage due to increased fragmentation or ablation. This survival bias would skew the observational dataset towards larger bolides with shallower entry angles, thereby reinforcing the observed relationship between trajectory geometry, object size, and back azimuth residuals. Further investigation with an expanded dataset will be necessary to confirm whether these correlations represent true physical phenomena or are artifacts of limited sampling. Most bolides analyzed have impact velocities below 20 km/s, with the exception of one event, 21 November 2009 (Figure 8c). While back azimuth residuals are relatively evenly distributed across all ranges (Figure 8d), the most significant scatter in residuals is observed for events with shallow entry angles.

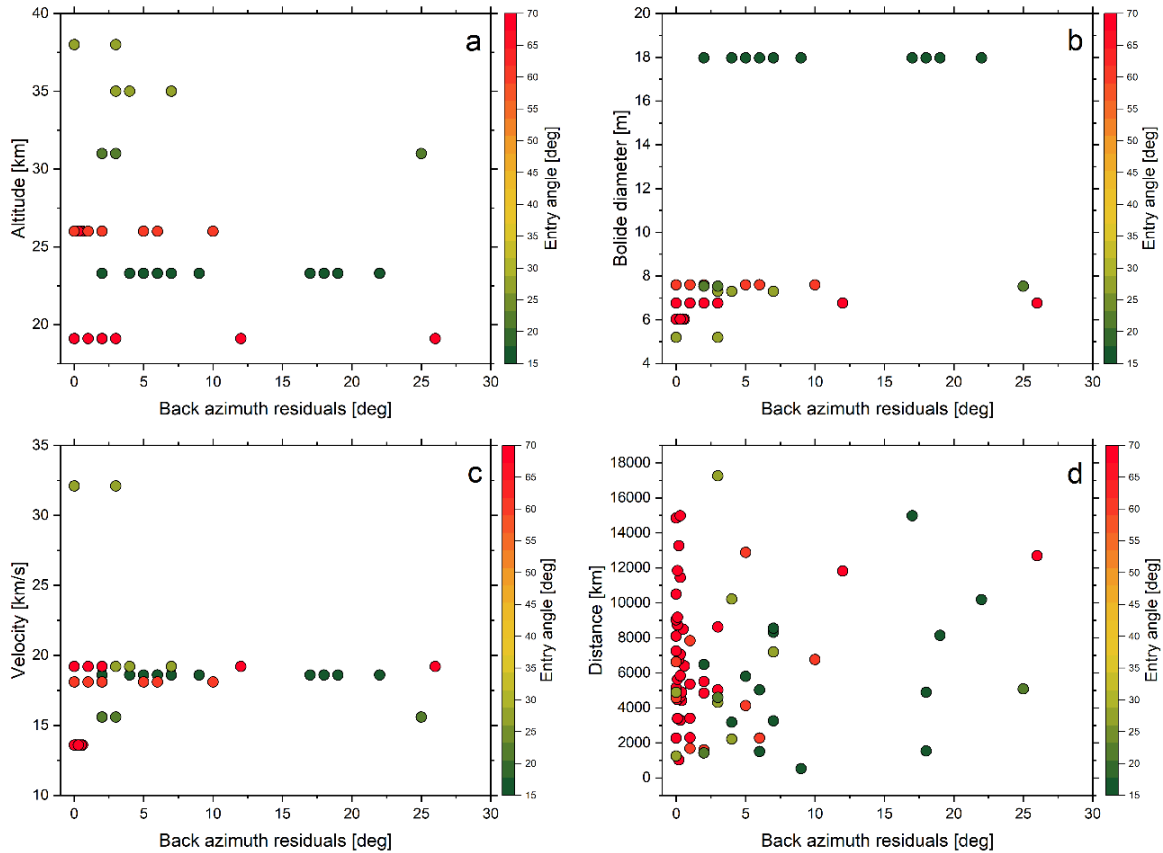

**Figure 8:** Panel plot illustrating the relationships between back azimuth residuals and various bolide quantities (a) peak brightness altitude, (b) object diameter, (c) impact velocity, and (d) source-station distance). Data points are color-coded according to the bolide entry angle, showing how variations in entry angle influence the observed deviations across different physical parameters. This visualization emphasizes that shallow entry angles tend to yield larger residuals, while steeper entries are associated with smaller deviations.

## 3.2 Synthetic Bolides

The outputs generated by BIBEX-M, including computed back azimuth deviations and distances for all entry angles considered here, are in the Supplementary Data available via an external repository (see Data Availability section). Figure 9a illustrates the ground track length as a function of entry angle. Bolides entering at very shallow angles exhibit long ground tracks, extending several hundred kilometers, whereas those with steep entry angles have much shorter ground tracks, covering only a few tens of kilometers. While this correlation is well-established, it is essential to further explore how trajectory geometry influences the apparent back azimuths. The representative results from the model are presented in Figures 9b and 9c. For a bolide with an entry angle of 30°, back azimuth residuals can reach up to 13.9° (mean 6.1°) at distances of 1000 km and 4.7° (mean 2.0°) at distances of up to 3000 km (see Table 2 and Figure 9b). When an entry angle is reduced to 10° (Figure 9c), the residuals reach up to 46.3° (mean 20.4°) at distances of less than 1000 km, persisting with residuals of up to 15.3° at 3000 km, and remaining non-negligible at greater distances. For example, the residuals are up to 7.9° with a mean of 3.9° at 5000 km and 5.0° with a mean 2.5° at 10,000 km. The results for the entire suite of simulations are included in the supplemental materials, and results for five representative entry angles are summarized in Table 2. It is important to reiterate that the residuals reported in Table 2 arise solely from the geometric effects associated with the bolide's trajectory (i.e., entry angle), without consideration of factors such as signal propagation, station characteristics, or signal processing, which may further modulate these deviations. In Table 2, not only the mean but also the median and minimum residual values are reported, which serve to characterize the plateau behavior. This multi-faceted approach enables the distinction between transient effects observed at short ranges and asymptotic values evident at larger distances, thereby offering a more comprehensive understanding of the geometric influences on back azimuth deviations.

The heatmap in Figure 10 shows the maximum back azimuth residuals binned in 1000 km, with the dynamic range of the colormap presented on a logarithmic scale to better visualize the variations.

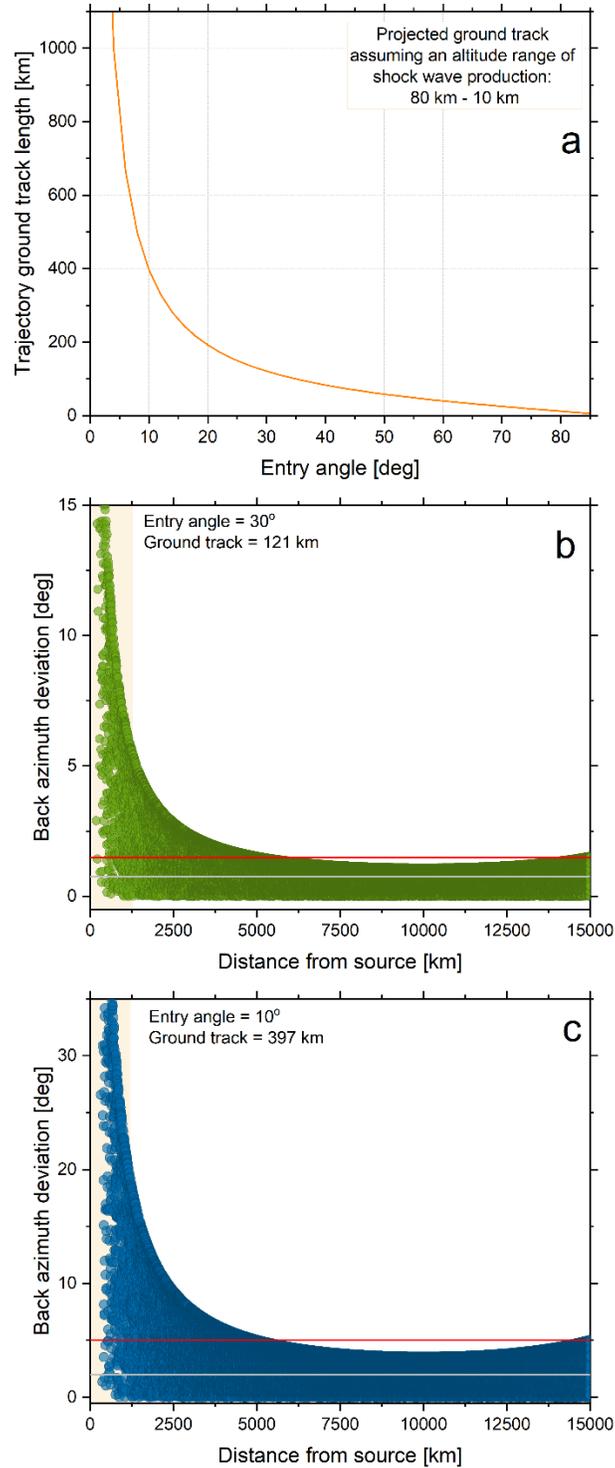

**Figure 9:** (a) Ground track length as a function of entry angle; back azimuth deviations (residuals) for synthetic bolides entering at an angle of (b) 30° and (c) 10°. The y-axis in panels (b) and (c) have been truncated for clarity; the full range of back azimuth residuals is provided in Table 2 and the supplemental materials.

**Table 2:** Results for five representative scenarios with entry angles of 10°, 30°, 45°, 60° and 75°. The resultant back azimuth residuals are listed in the last three columns (mean, median, and minimum). The results for all synthetic bolides across all entry angles are in the supplemental materials.

| Entry Angle [deg] | Distance Range | Mean Deviation [deg] | Median Deviation [deg] | Max Deviation [deg] |
|---|---|---|---|---|
| 10 | <500 km | 52.4 | 48.9 | 174.7 |
| 10 | 501-1000 km | 20.4 | 20.8 | 46.3 |
| 10 | 1001-1500 km | 12.0 | 12.9 | 22.8 |
| 10 | 1501-3000 km | 6.6 | 6.9 | 15.3 |
| 10 | 3001-5000 km | 3.9 | 4.3 | 7.9 |
| 10 | 5001-10000 km | 2.5 | 2.8 | 5.0 |
| 10 | 10001-12000 km | 2.3 | 2.6 | 3.8 |
| 30 | <500 km | 12.7 | 13.3 | 30.3 |
| 30 | 501-1000 km | 6.1 | 6.3 | 13.9 |
| 30 | 1001-1500 km | 3.7 | 4.0 | 6.9 |
| 30 | 1501-3000 km | 2.0 | 2.1 | 4.7 |
| 30 | 3001-5000 km | 1.2 | 1.3 | 2.4 |
| 30 | 5001-10000 km | 0.8 | 0.8 | 1.5 |
| 30 | 10001-12000 km | 0.7 | 0.8 | 1.1 |
| 45 | <500 km | 7.3 | 7.8 | 16.8 |
| 45 | 501-1000 km | 3.5 | 3.7 | 7.7 |
| 45 | 1001-1500 km | 2.1 | 2.4 | 4.0 |
| 45 | 1501-3000 km | 1.2 | 1.2 | 2.7 |
| 45 | 3001-5000 km | 0.7 | 0.8 | 1.4 |
| 45 | 5001-10000 km | 0.4 | 0.5 | 0.9 |
| 45 | 10001-12000 km | 0.4 | 0.5 | 0.7 |
| 60 | <500 km | 4.2 | 4.4 | 9.8 |
| 60 | 501-1000 km | 2.0 | 2.1 | 4.6 |
| 60 | 1001-1500 km | 1.2 | 1.4 | 2.3 |
| 60 | 1501-3000 km | 0.7 | 0.7 | 1.6 |
| 60 | 3001-5000 km | 0.4 | 0.4 | 0.8 |
| 60 | 5001-10000 km | 0.3 | 0.3 | 0.5 |
| 60 | 10001-12000 km | 0.2 | 0.3 | 0.4 |
| 75 | <500 km | 1.9 | 2.0 | 4.6 |
| 75 | 501-1000 km | 0.9 | 1.0 | 2.1 |
| 75 | 1001-1500 km | 0.6 | 0.6 | 1.1 |
| 75 | 1501-3000 km | 0.3 | 0.3 | 0.7 |
| 75 | 3001-5000 km | 0.2 | 0.2 | 0.4 |
| 75 | 5001-10000 km | 0.1 | 0.1 | 0.2 |
| 75 | 10001-12000 km | 0.1 | 0.1 | 0.2 |

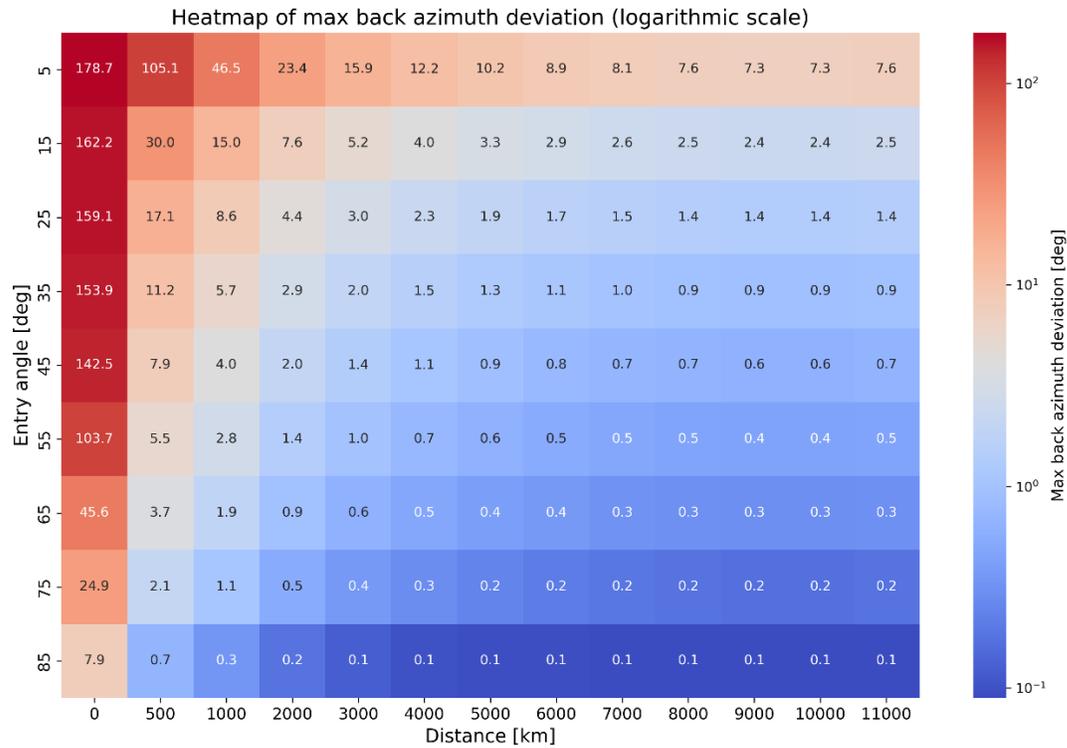

**Figure 10:** Heatmap depicting the maximum back azimuth deviations (residuals) binned in 1000 km increments. To better visualize the variations in back azimuth residuals, the dynamic range of colormap is displayed on a logarithmic scale.

While it might seem obvious that a moving source will exhibit azimuth deviations due solely to geometric factors, no previous study has rigorously quantified the upper bounds of these deviations. The significance of these simulations lies not only in systematically determining the maximum possible deviations as a function of both entry angle and distance, but also in illustrating the limitations of traditional point-source assumptions beyond certain ranges in infrasound analysis. These concrete numerical bounds are essential for advancing our understanding of atmospheric shock propagation and for improving the accuracy of event geolocation (e.g., Pilger et al., 2020; Silber, 2024b; Wilson et al., 2025). In an era where global monitoring networks increasingly detect high-energy events, ranging from bolides to reentry events and orbital debris, such rigorous quantification is indispensable. By bridging the methodological gap between conventional infrasound techniques and dynamic analyses required for moving sources, this work provides a critical foundation for the integration of geometric effects into current modeling practices, ultimately improving the interpretation and reliability of infrasound data across diverse applications.

The findings of this study indicate that treating a bolide as a point source in far-field scenarios may not always be valid. Instead, the bolide's entry angle and length of its propagation path must be carefully considered. For instance, shallow entry angles, particularly for highly energetic events, results in an extended shock wave path, leading to significant deviations from the point source approximation. This is evident from Pilger et al. (2015), who observed directional acoustic emissions from the Chelyabinsk event at IMS stations. The shallow entry angle of 16° (e.g., Peña-Asensio et al., 2022; Popova et al., 2013), resulted in an extended shock wave path, producing noticeable deviations. Similarly, a regional re-entry event, as exemplified by NASA's OSIRIS-REx sample return mission (Silber et al., 2024), further demonstrates that extended trajectories cannot be approximated as point sources (Silber and Bowman, 2025). In contrast, for bolides with steeper entry angles, the point-source approximation is more applicable, although shock altitude and the mode of shock production remain critical influences. For example, the Bering Sea (18 December 2018) and Indonesian (8 October 2009) events, which had entry angles of 69° and 67°, respectively, exhibited minimal back azimuth residuals, suggesting that for steep entries, the infrasound arrival direction is more affected by propagation effects than by geometric factors. The presence of infrasound waveguides at different altitudes can modulate acoustic propagation (e.g., Albert et al., 2023; Green and Nippress, 2019), suggesting that propagation modeling could help constrain shock source altitudes (Silber, 2024b; Silber and Brown, 2014). By establishing concrete numerical bounds on the maximum deviations attributable solely to the geometry of moving sources, this work bridges the gap between traditional infrasound methodologies and the dynamic analyses required for bolides and similar events.

## 4. Conclusions

Infrasound sensing is a powerful tool for detecting and characterizing bolides and assessing impact risks. Its global monitoring capabilities make it indispensable for planetary defense, as infrasound can contribute to refining influx rate and estimating bolide impact energy. The locations reported for bolide events typically correspond to the point of peak brightness (e.g., CNEOS database); however, apparent signal back azimuth (direction of arrival) observed in infrasound data does not always align with these reported locations. This discrepancy emphasizes the critical need to consider object trajectory geometry when interpreting infrasound data.

This study examined the effect of bolide trajectory geometry on apparent infrasound signal back azimuth variability, specifically focusing on how object entry angles affect these measurements independent of extrinsic factors such as propagation effects, station noise, and signal processing techniques. By isolating the purely geometric contributions using the proposed BIBEX-M framework, concrete numerical bounds on the maximum possible deviations were established.

The findings suggest that energetic bolides with shallow entry angles can produce significant deviations in apparent back azimuths. For example, bolides entering at 10° exhibited residuals averaging 20.4° and peaking as high as 46.3° at distances below 1000 km, with deviations of up to 5.0° persisting even at 10,000 km. In contrast, bolides with steep entry angles (>60°) show much smaller deviations, generally under 5° at 1000 km and diminishing to less than 1° beyond 5000 km.

These results not only challenge the traditional point-source assumption in infrasound analysis but also provide a robust baseline for understanding uncertainty in bolide geolocation at different source-to-station ranges. This work offers an integrated perspective that bridges the methodological gap between traditional infrasound techniques, typically designed for stationary sources, and the dynamic analyses required for moving sources, such as bolides, reentry events, and orbital debris.

Future studies with larger datasets and integrated propagation modeling should be carried out to further validate and refine these findings, ultimately improving both the detection and characterization of atmospheric impact events. Moreover, these results are highly relevant for reentry phenomena (e.g., space missions, orbital debris), illustrating the broader applicability of this framework.

These findings suggest that bolides with shallow entry angles have greater uncertainty in azimuthal measurements over long distances, making accurate trajectory predictions more challenging. Meanwhile, steeper entry bolides exhibit more predictable and smaller deviations, particularly beyond 5000 km. This understanding of azimuthal deviations based on entry angle can improve models for bolide characterization and advance the accuracy of bolide geolocation.

This study provides a framework for establishing uncertainty bounds for bolide location estimates based on the anticipated variability of back azimuths due to trajectory geometry, which could potentially aid in refining and improving detection, geolocation, and characterization. The relevance of this work extends beyond bolides to other events such as re-entries; a recent example being NASA's OSIRIS-REx sample return capsule re-entry. Future studies are needed to further corroborate these findings and refine geolocation methods, especially in cases where ground truth is limited.

**Acknowledgements:** The author thanks the two anonymous reviewers and the editor for their valuable comments that helped improve this paper. Sandia National Laboratories is a multi-mission laboratory managed and operated by National Technology and Engineering Solutions of Sandia, LLC (NTESS), a wholly owned subsidiary of Honeywell International Inc., for the U.S. Department of Energy's National Nuclear Security Administration (DOE/NNSA) under contract DE-NA0003525. This written work is authored by an employee of NTESS. The employee, not NTESS, owns the right, title, and interest in and to the written work and is responsible for its contents. Any subjective views or opinions that might be expressed in the written work do not necessarily represent the views of the U.S. Government. The publisher acknowledges that the U.S. Government retains a non-exclusive, paid-up, irrevocable, world-wide license to publish or reproduce the published form of this written work or allow others to do so, for U.S. Government purposes. The DOE will provide public access to results of federally sponsored research in accordance with the DOE Public Access Plan.

**Funding:** This work was supported by the Laboratory Directed Research and Development (LDRD) program at Sandia National Laboratories, a multimission laboratory managed and operated by National Technology and Engineering Solutions of Sandia, LLC., a wholly owned subsidiary of Honeywell International, Inc., for the U.S. Department of Energy's National Nuclear Security Administration under contract DE-NA0003525.

**Data availability:** The CNEOS database can be found at: https://cneos.jpl.nasa.gov/fireballs/. The outputs generated using the Bolide Infrasound Back-Azimuth EXplorer Model (BIBEX-M) as part of this work will be made available on Harvard Dataverse.

**Conflict of interest:** The author declares no conflict of interest.